\documentclass[12pt,preprint]{aastex}

\slugcomment{in press at PASP}

\shorttitle{A UBV Photometric Survey of the Kepler Field}
\shortauthors{Everett, Howell \& Kinemuchi}

\begin{document}

\title{A UBV Photometric Survey of the {\it Kepler} Field}

\author{Mark E. Everett\altaffilmark{1}}
\affil{National Optical Astronomy Observatory}
\affil{950 North Cherry Ave., Tucson, AZ 85719}

\author{Steve B. Howell and Karen Kinemuchi\altaffilmark{1,2}}
\affil{NASA-Ames Research Center}
\affil{Mail Stop 244-30, Moffett Field, CA 94035}

\altaffiltext{1}{Visiting Astronomer, Kitt Peak National Observatory,
  National Optical Astronomy Observatory, which is operated by the
  Association of Universities for Research in Astronomy (AURA) under
  cooperative agreement with the National Science Foundation.}
\altaffiltext{2}{Bay Area Environmental Research Institute}

\begin{abstract}
  We present the motivations for and methods we used to create a new
  ground-based photometric survey of the field targeted by the NASA {\it
    Kepler} Mission.  The survey contains magnitudes for 4414002 sources
  in one or more of the $UBV$ filters, including 1862902 sources
  detected in all three filters.  The typical completeness limit is
  $U\sim18.7$, $B\sim19.3$, and $V\sim19.1$ magnitudes, but varies by
  location.  The area covered is 191 square degrees and includes the
  areas on and between the 42 {\it Kepler} CCDs as well as additional
  areas around the perimeter of the {\it Kepler} field.  The major
  significance of this survey is our addition of $U$ to the optical
  bandpass coverage available in the Kepler Input Catalog, which was
  primarily limited to the redder SDSS griz and D51 filters.  The $U$
  coverage reveals a sample of the hottest sources in the field, many of
  which are not currently targeted by {\it Kepler}, but may be objects
  of astrophysical interest.
\end{abstract}


\section{Introduction}\label{sect:introduction}
The NASA {\it Kepler} Mission uses a 0.95m aperture space telescope
targeting $>165000$ targets in a 115 square degree field in order to
produce long-term light curves with the primary mission of detecting
and characterizing transiting exoplanets \citep{boruckietal10}.  In
addition to the exoplanetary mission, the high-precision, long-term
light curves have proved invaluable for studying many other astrophysical
phenomena \citep[e.g.,][]{balonaetal11, barthetal11, basrietal11,
  chaplinetal11, kawaleretal10}.  Some of these astrophysical studies
use stars selected on the basis of properties other than those
indicating suitability as transiting exoplanet targets.

We carried out our survey of the {\it Kepler} field in $UBV$ filters
primarily to create a resource for selecting new {\it Kepler} targets,
although the new bandpass coverage also provides information that can
help characterize the existing target list.  Previous to this survey,
the only similarly deep, optical survey covering the entire {\it Kepler}
field in optical bandpasses had been the effort to create the {\it
  Kepler} Input Catalog (hereafter KIC; \citet{brownetal11}) for
mission planning and target selection.  The KIC is mainly limited to
the SDSS $griz$ filters and gravity-sensitive $D51$ bandpass near
515~nm.  Our new survey in $UBV$ provides photometric information
across the Balmer Jump, and should be especially helpful in selecting new
blue targets like O and B stars, hot white dwarfs, hot subdwarfs,
planetary nebula central stars, cataclysmic variables, and AGN.  At
the same time, the {\it Kepler}-INT Survey, a collaboration of the
UVEX \citep{grootetal09} and IPHAS \citep{drewetal05} surveys, is
surveying the {\it Kepler} field using $Ugri$ and H$\alpha$ filters
\citep[see][]{greissetal12}.

In this paper we describe our new 191 square degree $UBV$ photometric
survey of the {\it Kepler} field.  In \S~\ref{sect:observations} we
describe our observations.  In \S~\ref{sect:datareduction} we describe
the image reductions, our methods to find instrumental magnitudes for
sources in each exposure (under conditions that were not always
photometric), and how we tied those magnitudes to the same scale using
existing photometry from the KIC.  In \S~\ref{sect:results} we present
some basic results from the survey describing the usefulness of a
photometric catalog produced by this survey as a resource for
selecting new {\it Kepler} targets.  In \S~\ref{sect:catalog} we
announce the public availability of the data.

\section{Observations}\label{sect:observations}

We observed the {\it Kepler} field using the NOAO Mosaic-1.1 Wide
Field Imager and the WIYN 0.9m telescope on Kitt Peak, Arizona over
the course of 5 nights between UT 2011 June 23-27.  The observing
conditions were partial moon illumination with a mixture of
photometric skies and light clouds.  The seeing FWHM ranged from
$1\farcs3-3\farcs4$ in $U$, $1\farcs2-2\farcs9$ in $B$, and
$1\farcs2-2\farcs5$ in $V$ with median seeing values of $1\farcs7$,
$1\farcs5$, and $1\farcs5$ in $U$, $B$, and $V$ respectively.  The
Mosaic-1.1 Wide Field Imager employs a mosaic of 8 $2048\times4096$
pixel, thinned, AR-coated e2v CCDs \citep[see][]{sawyeretal10}.  Each
CCD is read out through 2 amplifiers that read out $1024\times4096$
pixel subarrays.  On the 0.9m telescope, the CCD mosaic is arranged in
a $4\times2$ (North-South by East-West) pattern and spans a
$59\arcmin\times59\arcmin$ field-of-view with approximately
$35\arcsec$ wide gaps between the CCDs and a plate scale of
$0\farcs43$~pixel$^{-1}$.  The controllers provide 18-bit resolution
in counts with a well depth exceeding 200,000 electrons.  Details of
the Mosaic-1.1 imager can be found at the Kitt Peak National
Observatory web
site\footnote{http://www.noao.edu/kpno/mosaic/mosaic.html}.  We
observed using Johnson/Harris filters (Kitt Peak serial numbers k1001,
k1002, and k1003 for $U$, $B$, and $V$ 
respectively)\footnote{http://www.noao.edu/kpno/mosaic/filters/}.

\begin{figure}
\epsscale{0.66}
\plotone{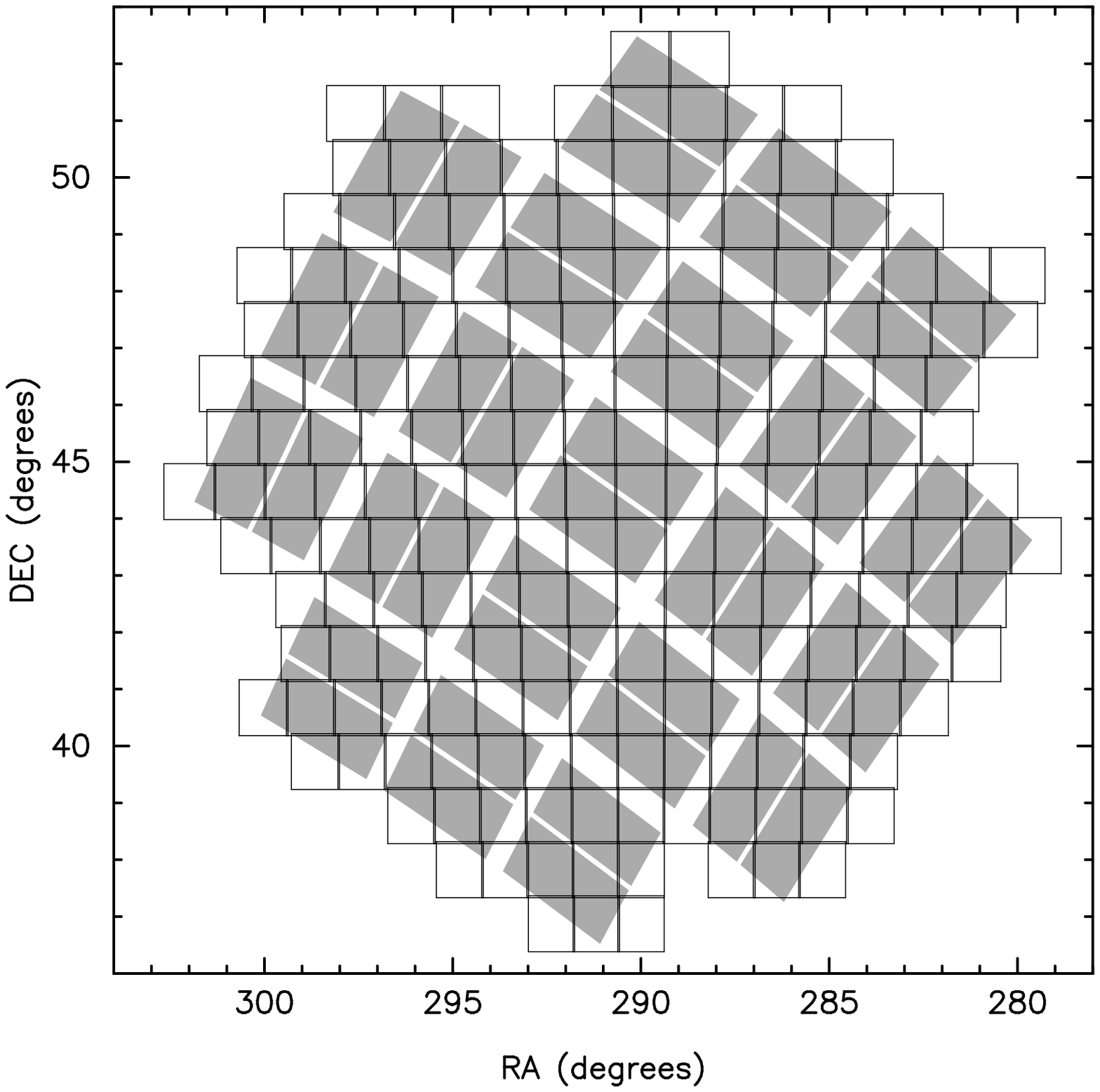}
\caption{
  The grid of 206 slightly-overlapping $59\arcmin\times59\arcmin$
  pointings in our 191 square degree survey are shown as boxes overlaid
  on the grey regions representing the 42 {\it Kepler} CCDs.
}
\label{fig:pointings}
\end{figure}

We selected a set of 206 pointings that covered the {\it Kepler} field
and surrounding areas in a grid pattern with $\sim1$~arcmin overlaps
between images at adjacent pointings.  The position of our pointings
and their relationship to the {\it Kepler} CCDs is shown in
Figure~\ref{fig:pointings}.  A few small areas at the edges of the
{\it Kepler} field, the gaps between the Mosaic-1.1 CCDs, and a few
areas masked out in our images account for the only areas within the
survey region lacking photometry.  These masks are defined to exclude
bad or saturated pixels, cosmic rays, areas within halos surrounding
the brightest stars, reflections in the optics that probably arise
from stars lying just outside the images, and satellite trails.  We
estimate the total area covered at 191 square degrees.

At each pointing, we took consecutive exposures in $U$, $B$ and $V$
with exposure times of 180, 40, and 40~seconds respectively before
moving to the next pointing.  We observed some pointings (usually in all 3
filters) more than once after inspecting the quality of each image for
poor seeing or lack of focus (the telescope focus drifted
significantly during some exposures).  Ultimately, only one exposure
per filter per pointing is used in the survey.  We tried whenever
possible to use images that were taken in close succession (within
$\sim5$~minutes of one another).  Our observations span an airmass
range of $1.0-1.6$.

We obtained bias frames and dome flats each night for data calibration
purposes.

\section{Data Reduction}\label{sect:datareduction}

\subsection{Pipeline Image Processing}

Our image reduction methods start with a NOAO-based reduction
pipeline that was developed for the Mosaic Camera and the Mayall 4-m
Telescope \citep{swaters-valdes07,valdes-swaters07}.
The 618 images (206 pointings with exposures in 3 filters at
each) are input to the pipeline along with bias and dome flat data.
The pipeline subtracts an overscan bias level from each image and a
residual bias pattern for each CCD based on bias frames taken during
the observing run.  It then flatfields each image using the dome
flats.  Because the dome flat field screen is not ideal, the flattened
images at this point have residual, large-scale non-flatness patterns.
To correct these, the pipeline constructs a master sky flat using all
dome flat corrected on-sky images taken during the run with deviant
pixels (e.g., stars) filtered out.  The sky flat is then applied to
each image.  The resulting product, known as ``InstCal'' images,
is the basis of our survey photometry.

Unfortunately, there appeared to be a time-variable component in the
2-D bias pattern that manifested itself as changes to a low spatial
frequency ``bounce'' or ``ringing'' pattern in the bias level across
the beginning of each row next to the output amplifiers.  Since this
pattern wasn't successfully subtracted from each individual image, it
remained in the dome flat corrected images.  From there, the same
pattern emerged in the sky flat fields, where, if not removed, would
have imposed an artificial pattern on fluxes along the edges of each
CCD.  In order to correct this problem, we chose to correct the sky
flats by fitting a single 1-D function to the average of all rows of
the flats in each readout region.  We fitted the flux of the flats,
$f$, for each readout region separately, parameterized using the
coefficients $a_1...a_6$:

\begin{eqnarray}
f(x) = a_1sin\left(\frac{x-a_2}{a_3}\right)\left(\frac{2}{2+e^{x/a_4}+e^{-x/a_4}}\right) + a_5 + a_6x
\end{eqnarray}  

\noindent
Here, $x$ is the column number of each readout region as measured from
the side with the amplifier.  This function has a form that is a
damped sine wave tilted to follow a line of constant slope.  A
correction to each sky flat is made by creating an image with values
along each row of the readout regions given by $f(x)/(a_5+a_6x)$.  We
corrected the sky flats by dividing them by these corrections.
Photometry of stars that are found on multiple overlapping images
confirms the success of this procedure.

The pipeline reduction also detects saturated pixels, cosmic rays, CCD
crosstalk features, and maps known bad pixels to create a bad pixel
mask for each image.  In addition, we manually inspected images for
artifacts such as halos surrounding the brightest stars, reflections,
and satellite trails.  We added these features to the bad pixel masks
to exclude the affected stars from our photometry.

Our astrometric plate solutions are also taken from the NOAO pipeline.
To find plate solutions, the pipeline locates USNO-B1.0 
Catalog\footnote{See catalog I/284 at
  http://vizier.u-strasbg.fr/viz-bin/VizieR} stars
\citep{monetetal03} in each image.  From the plate
solution, we find coordinates for the photocenter of each source.

\subsection{DAOPHOT Photometry}\label{sect:daophot}

Our first step toward getting photometry of all the sources in our
images is based on a recent version of DAOPHOT \citep{stetson87} that
is run separately on the exposure of each CCD.  Our use of DAOPHOT
consists of usual methods, and a minimum number of iterative steps
which seem to be sufficient at the stellar densities encountered.  The
first step in the process is to find objects in each image using the
``find'' routine.  Find detects the locations of flux peaks above a
certain threshold relative to the expected noise on each CCD.  The
detection threshold can be varied, as can the allowable bounds on the
``sharp'' and ``round'' criteria describing the shapes of objects at
each peak.  We selected detection parameters in an attempt to maximize the
detection of real sources and to minimize false detections.  This method
is fairly reliable in finding point sources and some galaxies with
bright centers, but tends to avoid detecting galaxies with more
diffuse light.  Following object detection, we run the aperture
photometry routine on each object, recording instrumental magnitudes
in a series of concentric circular apertures around each star.  In
this case, the sky flux is estimated and subtracted based on
measurements in an annulus centered on the stars and ranging between
radii of $8\farcs6$ and $15\farcs1$.  One aperture, with a radius of
$1\farcs3$, forms an initial estimate for the flux of each star.

We start DAOPHOT PSF-fitting photometry by picking a set of 40 optimal
stars in each CCD that appear relatively bright and uncrowded.  We use
the ``psf'' routine to construct a single PSF model based on fitting a
Penny model \citep{pennydickens86} to the 40 stars.  The PSF combines
elliptical Gaussian and Lorentz functions in which the Gaussian
component is rotated at an arbitrary angle but the Lorentz function is
constrained to be elongated only along the array axes.  Once the Penny
model parameters are fitted to match the selected stars, the remaining
component necessary to define the PSF is stored as a grid of look-up
values.  The resulting PSF can be scaled appropriately in flux and
subtracted to ``remove'' stars from the image.  We do this to remove
stars close to our set of 40 PSF modelling stars and repeat our model
fit to improve the PSF.  After this step, we adopt the PSF model as
our final solution and run the ``allstar'' routine to perform
PSF-fitting photometry on every star in the image.  In allstar, we fit
the PSF model to each star out to a radius of $2\farcs75$ and
simultaneously estimate the sky level at each star based on a
concentric annulus with inner and outer radii of $1\farcs3$ and
$8\farcs6$.

The final steps that we apply to the lists of instrumental magnitudes
for each exposure are aimed at removing stars that are false
detections of background noise or artifacts like diffraction spikes
from stars.  Since the PSF-fitting process produces various outputs
that correlate with false detections, we require that any sources
remaining in the photometry lists require no more than a maximum
number of iterations to be fit, that the $\chi^{2}_{\nu}$ of the PSF
fits not be too large, and that the ``sharp'' shape parameter not be
too high (as it would be for single-pixel features that are narrower
than the characteristic PSF of the image).  Lastly, even after
applying our automatic methods to remove false detections, we found 73
CCD images that contained an anomalously high number of faint
detections that could not be confirmed as valid objects upon visual
inspection.  These images are affected by unusually noisy readout
patterns, and the false detections were identified in these patterns.
For these images, we determined (by trial) a strict instrumental
magnitude cut-off and we removed all detections fainter than these
limits.

Once we had a final list of magnitudes, we cross-matched (by pixel
coordinates) the data taken in each filter for each CCD of each
pointing using the ``daomatch'' and ``daomaster'' programs, producing
a single set of stars for each pointing \citep[see the algorithm
  of][]{groth86}.  We applied our plate solutions to define a single
set of positions for each star.  Since adjacent pointings overlap, we
also have some stars that are common to between 2 and 4 neighboring
pointings.  We associate all of the photometry of these stars by
cross-matching the data by their celestial coordinates.  The
cross-matched sources are all found by searching within a radius of
$1\farcs0$ to allow for some error in the plate solutions in field
corners, although the mean separation between matched sources was
$0\farcs1$.

\subsection{Photometric Calibration}\label{sect:calibration}

The process of transforming our instrumental magnitudes measured in
each image to either a common magnitude scale or an established
photometric system is a complex task.  It is made more difficult due
to our desire to deliver this catalog to the public in a timely
manner, the considerable area of the survey, the occasional
non-photometric weather, and detailed calibration issues we would
likely encounter in transforming photometry obtained with the
Mosaic-1.1 Imager and Kitt Peak filters to
photometric systems established using different instruments.
Fortunately, optical photometry in the {\it Kepler} field already
exists with the KIC and we were able to use stars common to both our
data and the KIC to tie the magnitudes in our survey to a common
scale.  This is the most practical method to accomplish our primary
objective of obtaining a sample of hot, $U$-band enhanced sources for
use as future {\it Kepler} targets.

To correct our instrumental magnitudes to this common scale, we
compared the photometry obtained on each of our CCDs separately to
matching stars in the KIC, and to the same sources we observed
on adjacent, overlapping pointings.  We found a need to correct our
instrumental magnitudes using three correction factors:

\begin{eqnarray}
m(f,p,c) = m_{instr}(f,p,c) + C_{1}(f,p) + C_{2}(f,c) + C_{3}(f,p)
\end{eqnarray}

Here, $m$ is the corrected magnitude we wish to obtain, $m_{instr}$
is the instrumental magnitude we get from DAOPHOT as described in
\S~\ref{sect:daophot}, and $C_1$, $C_2$, and $C_3$ are separate
magnitude corrections that have distinct values that depend variously
on $f$, the filter used, $p$, the pointing observed, and $c$, the CCD
on which a star is found.  Although $C_1$ and $C_3$ have the same
dependencies, we calculate them separately.

Our first step in correcting instrumental magnitudes was to apply a
single offset, $C_1$, to the magnitudes measured in each exposure.
This correction simultaneously corrects for a number of effects
including first order atmospheric extinction, cloud extinction, and
exposure-to-exposure differences in an aperture correction.

To find $C_1$, we selected a sample of 33610 calibrator stars from the
KIC that are classified with $5500<T_{\rm eff}<6000$~K,
log(gravity)$>4.2$, and have $g<15$~magnitudes.  These stars were
chosen because they are bright, fall among a relatively narrow range
of colors, would be relatively unreddened, and are common in both
surveys.  We detected an average of 158 of these stars in each
exposure.  For each of these stars, we predicted a $B$ magnitude
using the transformation from \citet{jesteretal05}:

\begin{eqnarray}
B = g + 0.39(g-r) + 0.21
\end{eqnarray}

To predict $U$ and $V$ magnitudes for the same stars, we took the
empirical broadband colors from \citet{schmidt-kaler82} for G5 dwarfs
($U-B=0.20$; $B-V=0.68$) and G0 dwarfs ($U-B=0.06$; $B-V=0.58$) to
represent the predicted colors of stars with $T_{\rm eff}$ of 5500~K
and 6000~K respectively.  We predicted the colors of our calibrator
stars by linearly interpolating the colors in this range based on
values of $T_{\rm eff}$ in the KIC.  Initial values for $C_1$ for each
exposure could then be calculated as the median of the differences
between our instrumental and the predicted magnitudes (with $C_2$ and
$C_3$ momentarily set to zero).

The success of this procedure can be gauged by how well the magnitudes
agree for stars that are detected more than once in a given filter.
After our initial correction using $C_1$, the mean differences in
magnitudes between overlapping pointings were generally low (the mean
was 0.01 with standard deviation 0.02 magnitudes), but the range was 
$-0.15$ to 0.09 magnitudes.  Additionally, we found some offsets of
$\sim0.05$ magnitudes when we considered the mean differences in
magnitudes for individual pairs of overlapping CCDs.  Thus, we deemed
an additional correction $C_2$ to be necessary.  To find $C_2$, we
repeated the method we used to find $C_1$, but this time on a
CCD-by-CCD rather than exposure-by-exposure basis.  For each CCD, the
different values of $C_2$ calculated for each exposure showed more
scatter than did $C_1$, probably due to the smaller number of
comparison stars available in each CCD.  However, there was a similar
pattern in each exposure, so we defined a set of 24 initial $C_2$
values (for the 8 CCDs and 3 filters) to be the median of the
remaining corrections needed on each CCD after $C_1$ had been applied.
The range of $C_2$ values (comparing the most disparate CCDs) spanned
0.090 magnitudes in $U$, 0.082 magnitudes in $B$ and 0.048 magnitudes
in $V$.

Once we had initial values for both $C_1$ and $C_2$, we repeated the
process once to recalculate a final set of $C_1$ values with our
initial $C_2$ corrections applied.  We then recalculated final values
for $C_2$ with the final $C_1$ values applied.  At this point,
some of our exposures, which were scattered across the survey
area, still showed magnitude offsets relative to neighboring
pointings (as seen by systematic offsets in the magnitudes in
overlapping image edges).  The reason for these problematic exposures
is unknown, but because $\sim50$\% or more of the
adjacent exposures showed good agreement with one another, they can be
used to anchor the magnitude scale across the field in each filter.
We found a final set of corrections, $C_3$, to apply to 110 out of 206
$U$-band exposures, 91 out of 206 $B$-band exposures, and 65 out of
206 $V$-band exposures that reduces the remaining differences in
magnitude measured in adjacent pointings.  The values of $C_3$ were
defined as the minimum correction needed to bring the mean difference
in magnitudes of the deviant exposures to within 0.02 magnitudes of
the mean of their overlapping neighbors, or set to zero otherwise.
Standard deviations among the non-zero values of $C_3$ were 0.019,
0.015, 0.012 magnitudes for $U$, $B$, and $V$ filters respectively,
but absolute values of $C_3$ ranged as high as 0.061 magnitudes for
one exposure.

Finally, we associate each magnitude with an uncertainty arising from
both random noise characteristics and systematic errors.  We define
the magnitude uncertainties as the quadrature sum of an error assigned
by DAOPHOT, that incorporates the effects of noise and errors
associated with fitting the PSF and sky subtraction, and a mostly
systematic error that we find from our multiple exposures of the same
stars.  To determine the latter error component, we compare the mean
differences in magnitudes observed on different CCDs for pairs of
overlapping pointings.  We define this systematic error component to
be equal to $1/\sqrt{2}$ times the standard deviations of the magnitude 
differences, or 0.020 magnitudes in $U$, 0.023 magnitudes in $B$,
and 0.017 magnitudes in $V$.

\section{Results}\label{sect:results}

Our survey contains magnitudes for 4414002 sources in one or more of
the $UBV$ filters, including 1862902 sources detected in all three
filters.  This completeness in all three filters is primarily limited
by the $U$-band data, for which we have 1895173 detections.  In $B$ we
have 3089223 detections, and in $V$, our most complete bandpass, we
find 4363394 sources.  A few sources remain that are spurious
detections of background noise or features like diffraction spikes
from other stars.  These false detections could be essentially
eliminated from a sample of catalog entries by requiring detections in
two or more filters.

\begin{figure}
\plotone{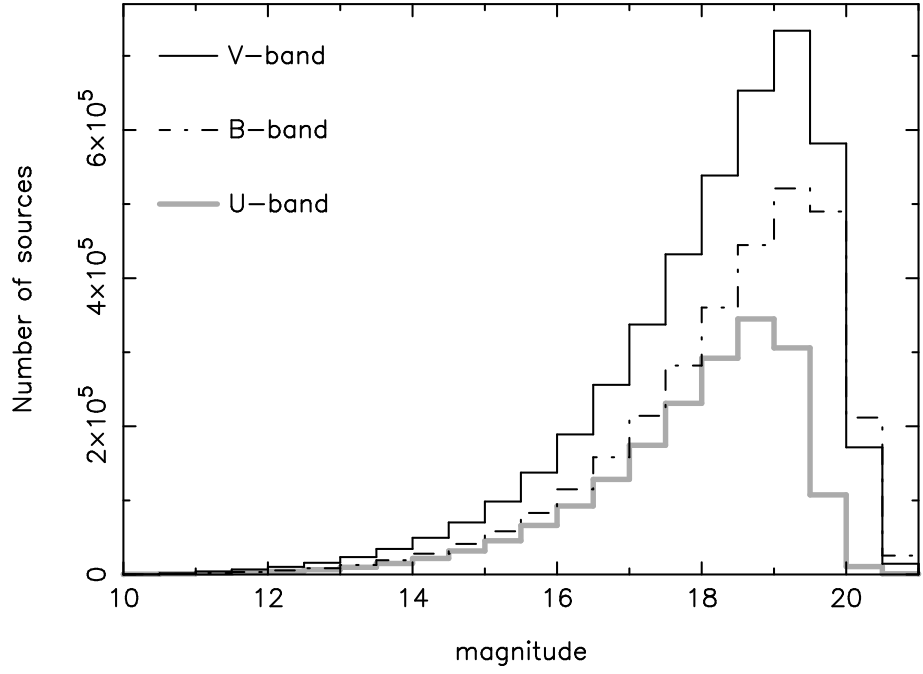}
\caption{
  A histogram of the number of detected sources as functions of $U$,
  $B$, and $V$ magnitudes.
}
\label{fig:mag_distribution}
\end{figure}

We plot the distribution of magnitudes in
Fig.~\ref{fig:mag_distribution}, which shows the wide dynamic range of
the survey.  The shape of the distribution is similar in each filter,
but shifted toward slightly fainter magnitudes in $B$ and brighter
magnitudes in $U$ when compared to $V$.  While there is variation in
our sensitivity from one pointing to another, we wished to estimate a
rough, survey-wide completeness limit, the faint end of our magnitude
range within which we expect to find almost all point sources in the
field that lie on our detectors.  To do this, we note that a
logarithmically-scaled plot of number of sources vs. magnitude
increases steadily with magnitude until some appreciable fraction of
sources ($\sim2$\%) remain undetected in some exposures (probably due
to relatively poor seeing).  The overall survey appears complete to
magnitudes $U\sim18.7$, $B\sim19.3$ and $V\sim19.1$.  Of course there
are fainter sources in the survey (to about 1 magnitude fainter).  The
bright limits in our exposures are sensitive to the observing
conditions, in particular the seeing.  We estimate the bright limit,
and its variation with location in our survey from the mean and
standard deviations of the magnitudes of the brightest star in each
exposure.  For $U$ we find the mean and standard deviation to be 10.12
and 0.34, for $B$ it is 10.60 and 0.33, and for $V$ it is 10.55 and
0.30.

\begin{figure}
\epsscale{0.5}
\plotone{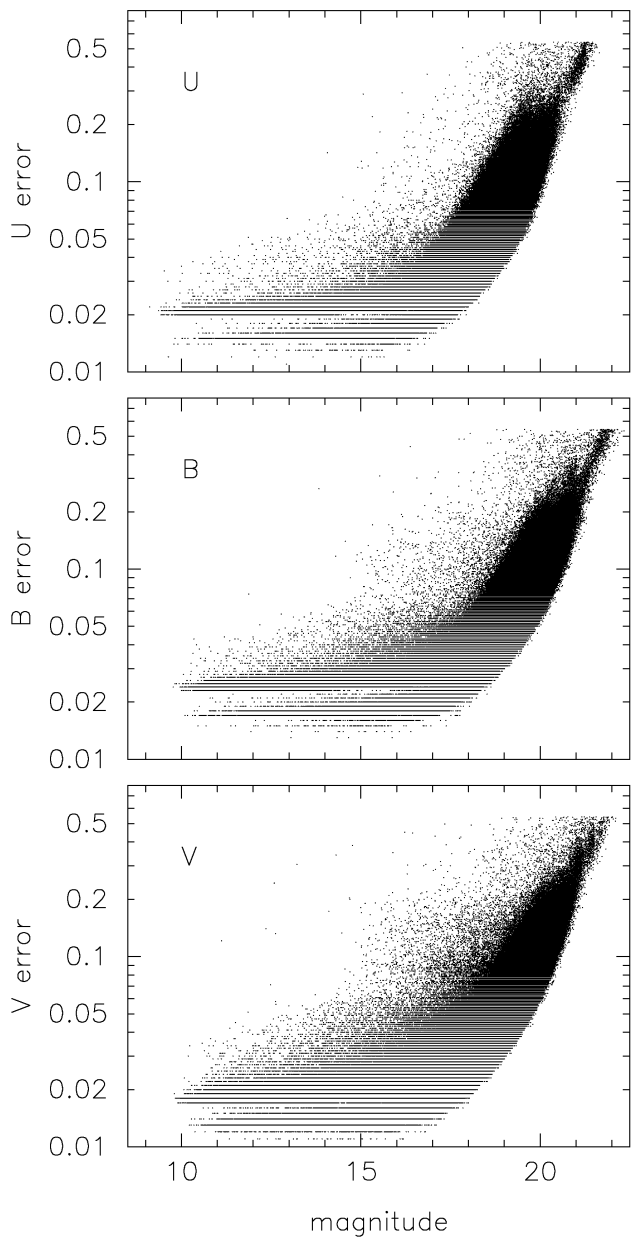}
\caption{
  The uncertainties in each magnitude plotted as a function of magnitude
  in each filter.  The distribution shows quantized values due to the limited
  precision quoted in our catalog.  At bright magnitudes, the errors converge
  toward lower limits reflecting the systematic errors adopted for each filter.
  Because some sources are observed up to four times each, there are lower
  uncertainties for some measurements, giving rise to multiple sequences of 
  points at low error values (see text for details).
}
\label{fig:mag_errors}
\end{figure}

In Fig.~\ref{fig:mag_errors} we plot the distribution of our
uncertainties in magnitude as a function of magnitude for each filter.
The uncertainties in each filter show a quantized distribution at low
values of error due to the limited precision quoted for our data.
With increasing brightness, the errors trend toward multiple lower
limits (two of a total of four in each filter are most easily seen).
The highest of these lower limits corresponds to the systematic
photometric uncertainties discussed in \S~\ref{sect:calibration}.  For
sources measured multiple times in overlapping pointings, our
uncertainty is lower, accounting for the secondary lower limits.  In
addition to the systematic error component, which was applied to
calculate each magnitude error, there are also exposure-to-exposure or
star-to-star differences in uncertainties that show up in the plot as
a broader scatter around the general trend versus magnitude.  These
errors can depend on the varying PSF in each image and source
crowding.  There are also uncertainties in our colors that can be
propagated from these magnitude uncertainties.  Color uncertainties
reach lower limits of 0.030 for $U-B$ and 0.029 for $B-V$ for sources
measured once in each filter.  Once again, the errors for bright
sources are mostly limited by the systematic errors discussed in
\S~\ref{sect:calibration}.

\begin{figure}
\epsscale{0.66}
\plotone{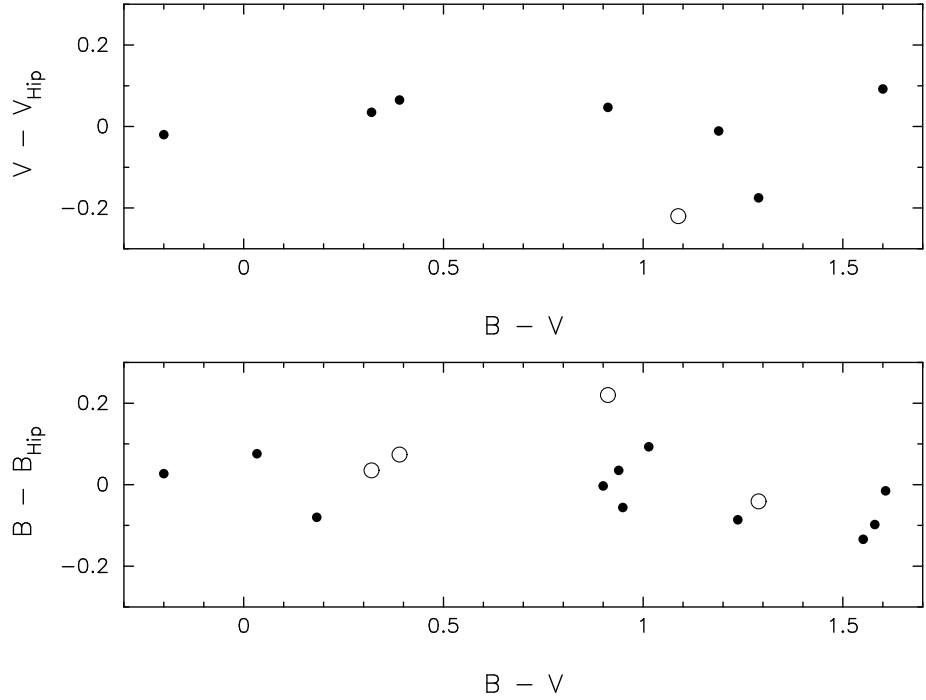}
\caption{
  A plot of the differences between our magnitudes and the corresponding
  Johnson magnitudes from the {\it Hipparcos} Catalogue as a function of
  the $B-V$ colors.  The top panel shows a comparison of $V$ magnitudes
  while the bottom panel shows a comparison of $B$ magnitudes.  The
  $B-V$ colors are taken from the {\it Hipparcos} Catalogue Johnson
  colors when available.  The $B-V$ colors of two stars are taken from
  our survey.  Stars having a {\it Hipparcos} Johnson magnitude fainter
  than 11.5 are shown as open circles.
}
\label{fig:us_vs_hipp}
\end{figure}

At this point we wish make a comparison of our magnitudes to those in
the {\it Hipparcos} Catalogue \citep{ESA97} because it is known to be
a source of well-calibrated photometry and was calibrated to match a
standard magnitude scale.  Unfortunately, the {\it Hipparcos}
Catalogue contains mostly sources brighter than those in our
survey.  Nevertheless, there are 19 sources in common to the two
catalogs that are not listed in the {\it Hipparcos} Catalogue as being
known or suspected variables according to the coarse variability or
HvarType flags.  In Fig.~\ref{fig:us_vs_hipp} we plot the differences
in $B$ and $V$ magnitudes between our survey and Johnson magnitudes in
the {\it Hipparcos} Catalogue as a function of their $B-V$ colors.
While most of the $B-V$ colors plotted are Johnson colors from {\it
  Hipparcos}, we use our own $B-V$ for the two stars lacking {\it
  Hipparcos} $B$ magnitudes.  There is one very discrepant datum of
$B-B_{\rm Hip}=-1.036$ for HIP95024, but with $B_{\rm Hip}=12.91$,
this star has a Hipparcos magnitude 0.87 magnitudes fainter than any
of the other stars we are comparing.  Magnitude errors increase
rapidly toward the faint limits of {\it Hipparcos} \citep{ESA97}, so
we leave this datum out of our analysis.  Other stars with Hipparcos
magnitudes fainter than 11.5 are shown using open symbols, but are
considered in our analysis.

A comparison of our magnitudes to {\it Hipparcos} Johnson magnitudes
reveals that there is an offset in the two magnitude scales of
$V-V_{\rm Hip} = -0.023$ with a standard deviation $\sigma_{\rm
  {\Delta}V} = 0.114$ and $B-B_{\rm Hip} = -0.008$ with a standard
deviation $\sigma_{\rm {\Delta}B} = 0.100$.  Thus, any differences
between the $B$ and $V$ magnitude scales of the two surveys are small.
The lack of a larger magnitude offset is perhaps surprising given the
considerable differences in the photometric methods used.

\begin{figure}
\plotone{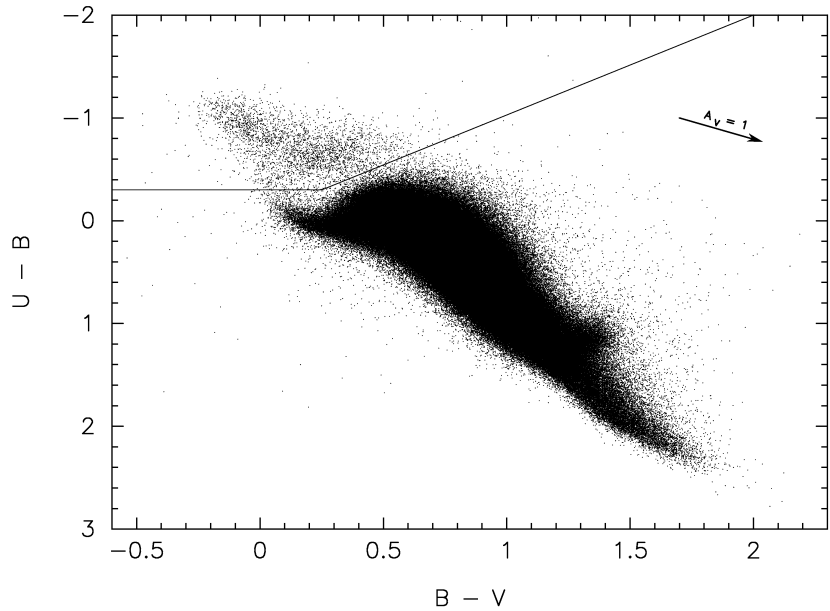}
\caption{
  A two-color diagram of the 1.8 million sources detected in 3 filters
  showing the dominant Main Sequence and Giant Branch populations along
  with some outlier colors.  The solid line segments define the upper
  limits in $U-B$ for a population of blue objects discussed in the
  text.  Here the color region is defined to be $U-B<-0.3$ (shown in the
  horizontal line segment) and $U-B<-0.971(B-V)-0.057$ (the diagonal
  line segment).
}
\label{fig:twocolordiagram}
\end{figure}

\begin{figure}
\plotone{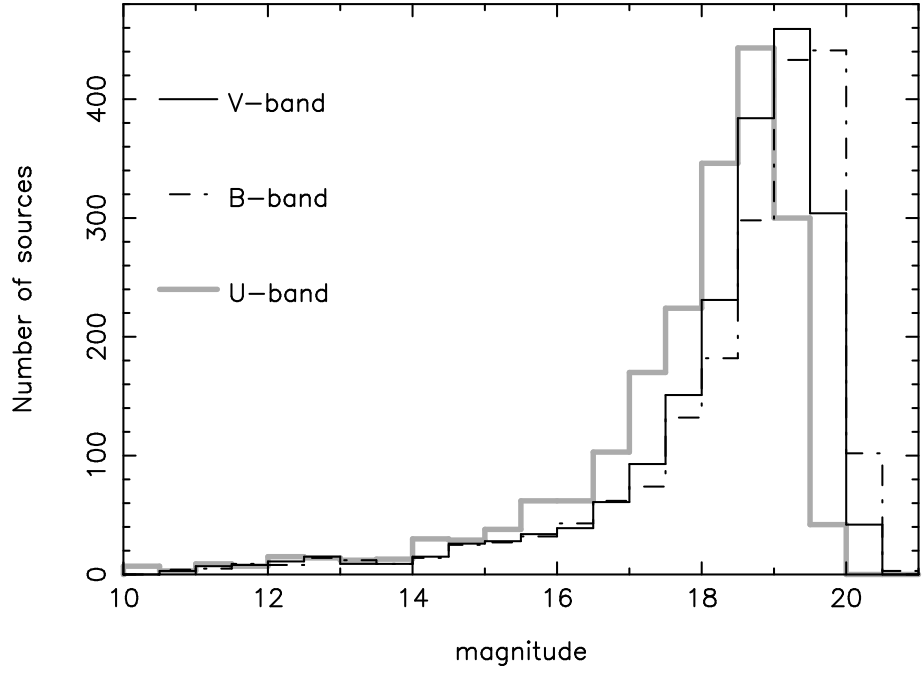}
\caption{
  A histogram of the number of the bluest sources that lie in the {\it
    Kepler} field as functions of $U$, $B$, and $V$ magnitudes.
}
\label{fig:bluemagnitudes}
\end{figure}

In Fig.~\ref{fig:twocolordiagram} we show a two-color diagram of
sources in the entire survey.  As expected, most sources lie along the
loci of the Main Sequence and Giant Branch.  To define a subsample of
the bluest sources in the survey, we select sources defined to satisfy
both of the following color conditions: $U-B<-0.3$ and
$U-B<-0.971(B-V)-0.057$.  This area of color space lies above the
lines in Fig.~\ref{fig:twocolordiagram} and contains 3092 sources.
The location of different stellar sources within this color region
depends on factors like effective temperature, reddening, and
composition.  In order to refine our sample to sources that are most
likely to be useful as new {\it Kepler} target stars, we define
rectangular regions for each {\it Kepler} CCD using coordinates
provided by the mission for one quarter of {\it Kepler} data.  We find
a subset of 1929 {\it Kepler} field sources among our blue sample.  A
histogram showing the magnitude distribution of these sources is given
in Fig.~\ref{fig:bluemagnitudes}.  While most of these sources are
fainter than the sample used for the mission's exoplanet search (ie.
fainter than $V\sim16$), they are sufficiently bright to allow
high-precision {\it Kepler} light curves and are good targets for
ground-based classification spectroscopy.  We anticipate that the
photometric catalog produced from our survey will serve as a rich
source of such targets.

\section{Data Availability}\label{sect:catalog}

The product of this survey is a publicly-available data catalog
available for downloading or use as part of a cross-matched,
multi-catalog, searchable database at the Multimission Archive at the
Space Telescope Science Institute
(MAST)\footnote{http://archive.stsci.edu/kepler/kepler\_fov/search.php
  and http://archive.stsci.edu/prepds/kplrubv/}.  The catalog includes
the data from this survey (positions, magnitudes and their
uncertainties) along with the products of other cross-matched
catalogs, including the KIC.  A detailed description of the catalog
contents may be found at the website.  In addition to the catalog, an
interactive tool is planned that will provide sections of our reduced
images to users for inspection.

\acknowledgments

We wish to thank the NOAO staff, including Frank Valdes, who helped us
to understand the data and worked to process our images through the
Mosaic image pipeline.  We are also grateful to the efforts of MAST for
their work in hosting our survey products.  This manuscript was also
improved by the helpful comments of a referee, who we wish to thank.

Funding for this research was provided by NASA {\it Kepler} grant
NNA04CK77G to SH awarded to NOAO. {\it Kepler} was selected as the
10th mission of the NASA Discovery Program.

The 0.9m telescope is operated by WIYN Inc. on behalf of a Consortium
of ten partner Universities and
Organizations\footnote{http://www.noao.edu/0.9m/general.html}. WIYN is
a joint partnership of the University of Wisconsin at Madison, Indiana
University, Yale University, and the National Optical Astronomical
Observatory.

\clearpage

\clearpage

\clearpage

\clearpage

\clearpage

\clearpage

\begin{thebibliography}{}
\bibitem[Balona et al.(2011)]{balonaetal11} Balona, L. A. et al. 2011,
  \mnras, 413, 2403
\bibitem[Barth et al.(2011)]{barthetal11} Barth, A. J. et al. 2011, \apj, 732, 121
\bibitem[Basri et al.(2011)]{basrietal11} Basri, G. et al. 2011, \aj,
  141, 20
\bibitem[Borucki et al.(2010)]{boruckietal10} Borucki, W. J. et al. 2010,
  Science, 327, 977
\bibitem[Brown et al.(2011)]{brownetal11} Brown, T. M., Latham, D. W.,
  Everett, M. E., \& Esquerdo, G. A. 2011, \aj, 142, 112
\bibitem[Chaplin et al.(2011)]{chaplinetal11} Chaplin, W. J. et al. 2011,
  Science, 332, 213
\bibitem[Drew et al.(2005)]{drewetal05} Drew, J. E. et al. 2005,
  \mnras, 362, 753
\bibitem[ESA(1997)]{ESA97}ESA, 1997, The Hipparcos and Tycho Catalogues, 
  ESA SP-1200
\bibitem[Greiss et al.(2012)]{greissetal12} Greiss, S. et al. 2012, 
  arXiv:1202.6333
\bibitem[Groot et al.(2009)]{grootetal09} Groot, P. J. et al. 2009,
  \mnras, 399, 323
\bibitem[Groth(1986)]{groth86} Groth, E. J. 1986, \aj, 91, 1244
\bibitem[Jester et al.(2005)]{jesteretal05} Jester, S. et al. 2005,
  \aj, 130, 873
\bibitem[Kawaler et al.(2010)]{kawaleretal10} Kawaler, S. D. et al. 2010, 
  409, 1509 
\bibitem[Monet et al.(2003)]{monetetal03} Monet, D. G. et al. 2003, \aj, 
  125, 984
\bibitem[Penny \& Dickens(1986)]{pennydickens86} Penny, A. J., \&
  Dickens, R. J.  1986, 220, 845
\bibitem[Sawyer et al.(2010)]{sawyeretal10} Sawyer, D. G., Daly, P. N., 
  Howell S. B., et al., 2010, SPIE, 7735, 111
\bibitem[Schmidt-Kaler(1982)]{schmidt-kaler82} Schmidt-Kaler, T. 1982
  in Landolt-B\"ornstein: Numerical Data and Functional Relationships in
  Science and Technology - New Series "Group 6 Astronomy and
  Astrophysics" Volume 2 Schaifers/Voigt: Astronomy and Astrophysics
  Stars and Star Clusters, XV
\bibitem[Stetson(1987)]{stetson87} Stetson, P. B. 1987, \pasp, 99, 191
\bibitem[Swaters \& Valdes(2007)]{swaters-valdes07} Swaters, R. A. \& Valdes, F. G.
  2007, in ASP Conf. Ser. 376, Astronomical Data Analysis Software and
  Systems XVI, eds. R. A. Shaw, F. Hill, \& D. J. Bell 
  (San Francisco: ASP), 269
\bibitem[Valdes \& Swaters(2007)]{valdes-swaters07} Valdes, F. G. \& Swaters, R. A.
  2007, in ASP Conf. Ser. 376, Astronomical Data Analysis Software and
  Systems XVI, eds. R. A. Shaw, F. Hill, \& D. J. Bell 
  (San Francisco: ASP), 273
\end{thebibliography}
\end{document}